\documentclass[prl,twocolumn,altaffillsymbol,showpacs,showkeys]{revtex4}

\usepackage{graphicx}% Include figure files
\usepackage{epsfig}% Include figure files
\usepackage{dcolumn}% Align table columns on decimal point
\usepackage{bm}% bold math
\usepackage{hyperref}% load hyper ref package 

\begin{document}

\preprint{APS/1}

\title{Momentum space properties from coordinate space electron density}

\author{ Manoj K. Harbola$^1$, Rajendra R. Zope$^{2,\dagger},$  
Anjali Kshirsagar$^3,$ and Rajeev K. Pathak$^{3,4}$}

\affiliation{%
$^1$
Department of Physics, Indian Institute of Technology, Kanpur 208 016, India
}

\affiliation{
$^2$
Department of Chemistry, George Washington University, Washington DC 20052, USA
}

\altaffiliation{
$\dagger$
Present mailing address : Theoretical Chemistry Section, Naval 
Research Laboratory, Washington DC 20375, USA.
}

\affiliation{
$^3$
Department of Physics, University of Pune, Pune 411007, India.
}

\affiliation{
$^4$
 The Abdus Salam International Center for Theoretical Physics,  Strada Costiera 11,
 34014, Trieste, Italy.
}

%\author{Manoj K. Harbola}
%\email{mkh@iitk.ac.in}
%\affiliation{%
% Department of Physics,\\
% Indian Institute of Technology,\\
% Kanpur 208 016, India
%}

%\author{Rajendra R. Zope }
%\email{rzope@alchemy.nrl.navy.mil}
%\affiliation{%
%Department of Chemistry,\\
%George Washington University,\\
%Washington DC 20052, USA
%}

%\author{Anjali Kshirsagar}
%\email{anjali@physics.unipune.ernet.in}
%% \homepage{http://www.Second.institution.edu/~Charlie.Author}
%\affiliation{
% Department of Physics, \\
%University of Pune, Pune 411007,
%India \\
%}

%\author{Rajeev K. Pathak}
%\email{pathak@physics.unipune.ernet.in}
%\affiliation{
% Department of Physics, \\
%University of Pune, Pune 411007,
%India \\
%}

%format is to be seen

\date{\today}

\begin{abstract}
Electron density and electron momentum density, while independently tractable
experimentally, bear no direct connection without going through the 
many-electron wave function. However, invoking a variant of the 
constrained-search formulation 
of density functional theory, we develop a general scheme  (valid for arbitrary
external 
potentials)  yielding decent momentum space properties, starting {\it {exclusively}} from the {\it {coordinate}}- space electron density.
Numerical illustration of the scheme is provided for the closed-shell
atomic systems He, Be and Ne and for $1s^1~2s^1$ singlet electronic
excited state for Helium by calculating the Compton profiles and the
$\langle p^n \rangle$ expectation values derived from given
coordinate space electron densities.
\end{abstract}

\pacs{31.15 Ew, 32.89 Cy}% PACS, the Physics and Astronomy
                             % Classification Scheme.
\keywords{Compton Profile, Electron Momentum Density, Electron Density}
%Use showkeys class option if keyword display desired

\maketitle{}
%\section{\label{sec:level1}First-level heading:\protect\\ The line
%break was forced \lowercase{via} \textbackslash\textbackslash}

For a quantum mechanical N-electron system such as an atom, molecule or a solid,
the coordinate space one-electron density $n(\vec r)$ is derived from the 
pertinent
configuration space many-electron (antisymmetric) wave function $\Psi 
({\vec r_1},
{\vec r_2}, {\vec r_3}, ..., {\vec r_N})$ through the marginal distribution
$n(\vec r) = N \int {\vert \Psi({\vec r},{\vec r_2},{\vec r_3},\ldots,
{\vec r_N})
\vert} ^2 d^3r_2 d^3 r_3 \ldots d^3 r_N$
(electron spins may also be included).
Analogously, the one-electron momentum density $\gamma (\vec p)$ is obtained
from the N-electron momentum space wave function
$\Phi ( {\vec p_1},{\vec p_2},{\vec p_3}, \ldots,{\vec p_N})$ via the reduction
$\gamma(\vec p) = N \int {\vert \Phi({\vec p},{\vec p_2},{\vec p_3},\ldots,
{\vec p_N}) \vert} ^2 d^3p_2 d^3 p_3  \ldots d^3 p_N.$ 
The wave functions $\Psi$ and $\Phi$ in the complementary 
%coordinate and momentum
spaces are connected by a many-particle Fourier-Dirac transformation~:
\begin{eqnarray}
\Phi({\vec p_1},{\vec p_2},\ldots,{\vec p_N}) & = & 
{\frac{1}{(2 \pi)^{3N/2}}}
\int \Psi({\vec r_1},{\vec r_2},\ldots,{\vec r_N}) \nonumber \\
 & & \times ~  e^{i {{\sum_{j=1}}^N} {\vec p_j}.{\vec r_j}} d^3 r_1d^3r_2 \ldots d^3 r_N. \nonumber \\
 & & 
\label{eq1}
\end{eqnarray}
\noindent 
(Hartree atomic units, {\it viz.} $\vert e \vert = 1, m_e = 1, \hbar = 1$ 
have been used throughout).  
%Scaled by 1/N, both the densities are 
%single-particle probability densities in the respective spaces.
Experimentally, the coordinate-space density is tractable through a coherent and
elastic x-ray scattering process \cite{march} where the scattered intensity is
directly proportional to ${\vert f({\vec k})\vert }^2$, $f( {\vec k})$ being the
form factor which is the Fourier transform of $n({\vec r})$.
On the other hand, the electron momentum density $\gamma ({\vec p})$ manifests
itself more directly in terms of the (directional) Compton profile $J(q)$ 
\cite{wil} obtained in an inelastic high energy (X-ray or $\gamma$-ray) 
Compton scattering process~:
 
\begin{equation}
J(q) =  \int_{-\infty}^{\infty} dp_x \int_{-\infty}^{\infty} dp_y
\gamma(p_x, p_y, q).  \label{eq2}
\end{equation}
 
For atomic and molecular systems in gaseous state, a spherically symmetric
Compton profile results from the corresponding spherically averaged electron 
momentum density
$\gamma_{sph} (p) = \frac{1}{4\pi} \int \gamma (\vec {p}) d\Omega_{\hat{p}}$,
whence
 
\begin{equation}
{J_{sph} (q)}_{q>0} = 2\pi \int_{q}^{\infty} \gamma_{sph} (p)~ p~ dp, 
\label{eq3}
\end{equation}                                         
\noindent
which is a monotonic decreasing function of $q$. Theoretically, the expressions in
Eqs. (\ref{eq2}) and (\ref{eq3}) are essentially the impulse approximation (IA)
profiles \cite{PT}. Equation (\ref{eq3}) readily leads to an inverse relation
$\gamma (p) = - \frac {1}{2 \pi p}  \frac{dJ(p)}{dp}$, where here and 
henceforth, the subscript ``sph'' will be understood.
 
It must be emphasized here that the mappings $ n \longrightarrow \Psi$ and 
$\gamma
\longrightarrow \Phi$ are both in general one-many and although for the 
ground state the former is unique \cite{HK}, the explicit prescription for 
the map
is unknown. Hence while there exists a
Fourier connection between $\Psi \longleftrightarrow \Phi$, no such direct 
relation
is possible between the densities $n(\vec{r})$ and $\gamma (\vec{p})$ in
the two complementary  spaces.
On the basis of quasi-classical phase-space considerations (akin to the
Thomas-Fermi theory), there exists a procedure due to Burkhardt \cite{bur}, 
K\`{o}nya \cite{kon}, and Coulson and March \cite{CM}, called the BKCM scheme,
to estimate electron momentum density, given its position-space counterpart 
\cite{PG}.  However, this method is marred by its artifacts of a divergent
$\gamma (0)$ and a finite cut-off for $\gamma(p)$.           
Incidentally, the so termed ``Wigner function'' \cite{wig,hil,KN}
cannot represent a joint probability
in phase space for not being strictly positive semidefinite.
In their phase-space approach to density functional theory (DFT) \cite{HK},
Parr {\it et al.} \cite{parr} prescribed a phase-space entropy maximization, 
imposing a given coordinate density {\it and} a given kinetic energy density 
(at each point $\vec{r}$) as constraints. This enabled them to obtain a 
positive semidefinite phase-space distribution through which momentum density
could be extracted.

Let us, however, pose a question : Given {\it exclusively} the electron density in
coordinate space as a starting point ({\it {and no other information}}), could one 
estimate
the quantum-mechanical electron momentum density (and hence the Compton
profile)? 

It is the spirit underlying this letter to demonstrate an affirmative
answer to the above question, within the density functional theory 
pertinent features of which have been 
as highlighted below.

In their exciting work, Zhao and Parr (ZP) \cite{ZP} developed a novel
method to obtain the Kohn-Sham orbitals for a given co-ordinate
space density. Their method is based on Levy's constrained search 
approach \cite{Levy} 
that generalizes the Hohenberg - Kohn formulation of DFT. 
Constrained search approach 
which obtains the Kohn-Sham ``wave function'',
a single Slater determinant $\Psi^D$ formed out of the lowest
occupied orbitals of a local potential, by minimizing the non-interacting
kinetic energy

\vspace{-0.2in}
 
\begin{equation}
T_s [n] = \min_{\Psi^D} <\Psi^D | \hat{T} | \Psi^D>, \label{eq4}
\end{equation}
 
\vspace{-0.1in}

\noindent
where $\Psi^D \Longrightarrow n(\vec{r}),$ the given density; $\hat{T}$ is 
the N-electron kinetic energy
operator and $\Psi^D$ is an antisymmetric, normalized (hence $L^2$) N-electron
wave function of independent electrons.
ZP accomplished the search on the right side of Eq. (\ref{eq4})
through variation of the orbitals $ \{ \psi_i (\vec{r}) \}_{i=1}^N$ yielding a
density
$n(\vec{r})$. This density would equal the given density $n_0(\vec{r})$
at every
point $\vec{r}$, via the minimization of the positive semidefinite functional

\vspace{-0.2in}

\begin{equation}
C \equiv \int \frac{[n(\vec{r}) - n_0(\vec{r})] [n(\vec{r'}) - n_0(\vec{r'})]}
{2 \mid \vec{r} - \vec{r'} \mid} \; \; \; \; d^3 r d^3 r', \nonumber
\end{equation}

\vspace{-0.1in}

\noindent
whose minimum value zero would be reached {\it iff} $n(\vec{r}) = n_0 (\vec{r})
\; \forall \vec{r}$. 
The minimization $T_s[n] + \lambda C$ with respect to the orbitals $\{ \psi_i \}$
gives a set of Kohn-Sham like equations~:

\begin{equation}
\left ( - \frac{\nabla^2}{2} + \lambda \tilde{v} (\vec{r})  -Z/r \right ) \psi_i
(\vec{r}) =
\epsilon_i \psi_i (\vec{r}), 
 (i = 1,\ldots,N) \nonumber \label{eq8}
\end{equation}

\noindent
where the sum 
$\lambda \tilde{v} (\vec{r}) - Z/r \equiv \lambda v(\vec{r}) = 
 \lambda  \int \frac{[n(\vec{r'}) - n_0(\vec{r'})]}
{\mid \vec{r} - \vec{r'} \mid}\;  d^3 r'$ in the limit $\lambda 
\rightarrow \infty, C \rightarrow 0$, yet the product $\lambda C$ remaining 
finite, 
gives the effective Kohn-Sham potential \cite{ZP}.

Morrison and Zhao \cite{mor} applied this method to atoms while 
Ingamells and Handy \cite{ing} extended the work to molecular systems. 
Recently, Harbola \cite{har} observed that the Zhao-Parr procedure
could also be applied to obtain Kohn-Sham orbitals for an excited-state 
density, thereby demonstrating the general applicability of the method to 
ground- as well as excited-states.

%It is these Kohn - Sham orbitals, for sufficiently large values of $\lambda$ as
%in Eq.(\ref{eq8}) which will be employed here to effect the desired $n(\vec{r})
%\longrightarrow \gamma(\vec{p})$ ``transformation''. In a nutshell,
Our scheme to effect $n(\vec{r}) \rightarrow \gamma(\vec{p})$ ``transformation'' is~:
Start from a given density $n_0(\vec{r}) \Rightarrow $
Obtain the Kohn - Sham orbitals $\psi_i(\vec{r})$ via the Zhao - Parr
prescription $\sum_{i=1}^{occ} {\mid \psi_i(\vec{r}) \mid}^2 = n(\vec{r})
\Rightarrow $
Fourier transform $\psi_i(\vec{r}) \longrightarrow \phi_i(\vec{p})
\Rightarrow $
Obtain $\gamma(\vec{p}) = \sum_{i=1}^{occ} {\mid \phi_i(\vec{p})
\mid}^2; $ hence the Compton profile $J(q)$ and other momentum expectation 
values.
This procedure thus starts from only a given $n_0(\vec{r})$ and estimates 
$J(q)$ and $\langle p^n \rangle$ values.
We illustrate this ``$\vec{r}$-density to momentum-space properties'' (RDMP) 
scheme for the following atomic systems :

\begin{table}[h!] 
\begin{ruledtabular}
\begin{tabular}{lccl}
$He$& $1s^2$& $^1S$& (ground state configuration)\\
$He$& $1s^1 \, 2s^1$& $^1S$& (singlet lowest excited state configuration)\\
$Be$& $1s^2 \, 2s^2$ & $^1S$& (ground state configuration)\\
$Ne$& $1s^2 \, 2s^2 \, 2p^6$ & $^1S$& (ground state configuration)
\end{tabular}
\end{ruledtabular}
\end{table}

%\vspace{0.2in}
 
Accurate coordinate space densities employed as starting points for ground 
state  are due to Koga  {\it et al.} \cite{kog} for He, Esquivel and Bunge
\cite{esq} and  Bunge and Esquivel \cite{bun} for Be and Ne respectively
while the Coolidge and James {\cite{coo}} 
density was used for He excited state.
The value of $\lambda$ (cf. eq. \ref{eq8} ) was set to $\sim$5000,
leading to a sufficiently accurate self-consistent Kohn-Sham potential
converged to five places.

\begin{figure}

\epsfxsize=3.0in
\epsfysize=3.0in
{\epsfbox{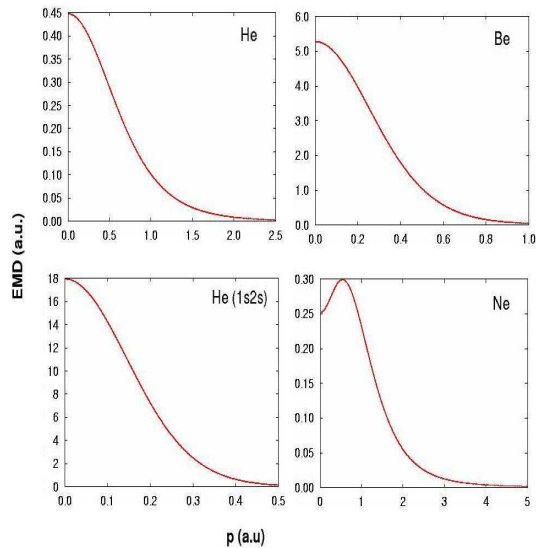}}
\vspace{0.3in}
\caption{Electron Momentum Density obtained within the present scheme }
\label{fig1} 

\end{figure}

A collage of the $\gamma(\vec{p})$ derived from $n(\vec{r})$ under the present
scheme is depicted in Fig. \ref{fig1}. While all the $\gamma$-plots compare 
extremely well with their HF-counterparts (not shown), the features of 
nonmonotonicity of $\gamma(\vec{p})$ for Neon is also reproduced (cf. Ref. 
\cite{GCP}; Fig. 1), in conformity with the fact that atoms with their
ultimate $p$-shells doubly occupied or more evince such nonmonotonicity \cite{GCP}.

The Hartree-Fock data (zeroth-order correlated) for the wave-function are given
for comparison and as a datum. Particularly striking is the fact that for
excited $He$, the electron momentum density (EMD) is overwhelmingly preponderant
around low momentum values along with a very sharp asymptotic fall-off. 
%Thus there is a
%major redistribution of momentum density.

%\vspace{-0.1in}
 
%\begin{table}[h!] 
\begin{table} 
%\caption{\label{tab:tI}Compton Profiles for Ground State of helium}
\caption{\label{tab:tI}Compton Profiles for helium}
\begin{ruledtabular}
\begin{tabular}{ccccc}
 & \multicolumn{3}{c}{Ground State}& {First Excited State}\\
$q$& Present Work & {HF}$^a$ &
{Experiment}$^b$ & Present Work \\ 
%$q$& Present Work & {HF}\footnote{Reference \cite{cle}} & 
%{Experiment}\footnote{Reference \cite{eis}}& Present Work \\ 
\hline
0.00 &1.075&1.070 &1.071 $\pm$ 1.5\% & 2.947\\
0.20 &1.020&1.017 &1.019&  1.464\\
0.40 &0.879&0.878 & -&  0.465\\
0.60 &0.700&0.700 &0.705&  0.337\\
0.80 &0.525&0.527 & -&  0.287\\
1.00 &0.380&0.382 &0.388&  0.233\\
1.50 &0.159&0.160 & -&  0.124\\
2.00 &0.068&0.068 &0.069&  0.059\\
2.50 &0.031&0.031 &0.030 $\pm$ 15\%&  0.027\\
3.00 &0.015&0.015 &0.013&  0.012\\
3.50 &0.008& -  & -&  0.006\\
4.00 &0.004& -  & -&  0.003\\
5.00 &0.001& -  & -&  0.001\\
%6.00 &0.000& -  & - & - 
\end{tabular}
\end{ruledtabular}

%\vspace{-0.2in}
$^a$Reference \cite{cle}; $^b$Reference \cite{eis}
\end{table}

\vspace{0.1in}

Tables \ref{tab:tI} through \ref{tab:tIV} give the CP's via the 
RDMP scheme with their other
theoretical and experimental comparators. 
Each CP is normalized as $ \int_{0}^{\infty} J(q) dq = N/2$, 
as is customary. There is a remarkable agreement between
the CP's derived from coordinate space atomic electron densities within
the impulse approximation and the
Hartree - Fock (HF) or experimental CP's. From Table \ref{tab:tI},
%and \ref{tab:tII}, 
the considerable redistribution of electron momentum density of He 
excited state in comparison with its ground state counterpart is evident.
Excitation brings in a slower decay in the coordinate space which is, by Fourier
reciprocity, mapped on to a peak Compton profile value.
As the transition $1s^1 \, 2s^1 \rightarrow 1s^2$ is dipole-forbidden, this
singlet He-excited state is  long-lived (life time $\sim \frac{1}{10} sec$);
while  there have been no Compton profiles (CP's) reported for this system,
the present scheme accomplishes this.

For Be and Ne (Tables \ref{tab:tIII} and \ref{tab:tIV}) ground states, the 
CP's via the present schemes give better overall agreement with their
accurate correlated counterparts than do the HF-CP's. This is indicative of the correlation piece picked up by the Zhao-Parr scheme.
 
\begin{table} 
\caption{\label{tab:tIII}Compton Profiles for Ground State of Beryllium}
\begin{ruledtabular}
\begin{tabular}{cccc}
$q$& Present Work & HF$^a$ &
Correlated$^b$\\
%$q$& Present Work & HF\footnote{Reference \cite{cle}}& 
%Correlated\footnote{Reference \cite{tri}}\\ 
\hline
 0.00& 3.061& 3.159 & 2.953 \\
 0.30& 1.958& 1.950 & 1.936 \\
 0.50& 1.068& 1.032 & 1.098 \\
 0.70& 0.621& 0.600 & 0.658 \\
 0.80& 0.516& 0.503 &  - \\
 1.00& 0.413& 0.409 & 0.432 \\
 1.50& 0.310& 0.309 & 0.312 \\
 2.00& 0.224& 0.224 & 0.224 \\
 2.50& 0.153& 0.153&  - \\
 3.00& 0.102& 0.102&  0.102 \\
 3.50& 0.068& 0.068&  - \\
 4.00& 0.045& 0.045&  0.045 \\
 5.00& 0.021& 0.021&  0.020 \\
 6.00& 0.010& 0.010&  - \\
 7.00& 0.005& 0.005&  0.005 \\
 8.00& 0.003& 0.003&  - \\
% 9.00& 0.002& 0.002&  - \\
%10.00& 0.001& 0.001& -  
\end{tabular}
\end{ruledtabular}

%\vspace{-0.2in}
$^a$Reference \cite{cle}; $^b$Reference \cite{tri}
\end{table}

\vspace{0.1in}
\begin{table} 
\caption{\label{tab:tIV}Compton Profiles Ground State of Neon}
\begin{ruledtabular}
\begin{tabular}{cccc}
$q$& Present Work & HF & Experiment\footnote{References
\cite{TL} - \cite{eis1}}\\ 
\hline
0.00& 2.748&2.727 &2.762 \\
0.20& 2.716&2.696 &2.738 \\
0.40& 2.610&2.593 &2.630 \\
0.60& 2.423&2.413 &2.427 \\
0.80& 2.170&2.168 &2.162 \\
1.00& 1.883&1.889 &1.859 \\
1.50& 1.216&1.228 & - \\
2.00& 0.764&0.771 &0.765 \\
2.50& 0.499&0.501 &0.501 \\
3.00& 0.346&0.346 &0.359 \\
3.50& 0.254&0.253 &0.277 \\
4.00& 0.195&0.194 &0.210 \\
5.00& 0.125&0.125 &0.126 \\
6.00& 0.085&   - & -\\
7.00& 0.060&   - & -\\
8.00& 0.044&   - & -
\end{tabular}
\end{ruledtabular}
\end{table}

To gauge the overall quality of the electron momentum density ``derived'' from
the coordinate space density, we have computed the $\langle p^n \rangle$ values
under the RDMP scheme and have compared with those derived from accurate,
correlated atomic wave functions \cite{sar}, as well as from the near 
Hartree-Fock
wave functions \cite{cle}. 
Table \ref{tab:tV} illustrates that the present RDMP scheme successfully
obtains the
$\langle p^n \rangle$ values (for $n =$ -2, -1, 1, 2, 3 and 4) 
in agreement with both their HF and correlated counter parts.
For the case of He excited state,
preponderance of electron momentum density around very small as well as 
large $p$ values is conspicuous.  
  
\begin{table*} 
\caption{\label{tab:tV}Various expectation values $ \langle p^n \rangle$
for $n$ = -2, -1, 1, 2, 3 and 4}
\begin{ruledtabular}
\begin{tabular}{lccccccccccccc}
System&\multicolumn{3}{c}{He(1s$^2$)}&\multicolumn{3}{c}{He(1s$^1$~2s$^1$)}
&\multicolumn{3}{c}{Be}& &\multicolumn{3}{c}{Ne}\\
Property&PW&HF$^a$ &Corr$^b$& &PW& &PW&HF$^a$&Corr$^c$ & &PW&HF$^a$&Corr$^{d}$\\
%Property&PW&HF\footnote{Reference \cite{cle}}&Corr\footnote{Reference \cite{sar}}& 
%&PW& &PW&HF$^b$&Corr\footnote{Reference \cite{tri}} 
%& &PW&HF$^b$&Corr$^a \footnote{Reference \cite{tri1}}$\\
$\langle p^{-2} \rangle$&4.132&4.089&4.101& &43.047& &23.449&25.291&21.939& &5.583&5.480&5.553\\
$\langle p^{-1} \rangle$&2.149&2.141&2.139& &5.889& &6.122&6.318&5.909& &5.497&5.456&5.478\\
$\langle p^1 \rangle$&2.797&2.799&2.814& &2.036& &7.468&7.434&7.533& &35.156&35.196&35.241\\
$\langle p^2 \rangle$&5.734&5.723&5.805& &4.318& &29.183&29.146&29.333& &257.183&257.09&257.751\\
$\langle p^3 \rangle$&18.11&17.99&18.40& &63.86& &185.55&185.59&186.35& &3583.33&3584.3&3591.5\\
$\langle p^4 \rangle$&106.6&105.7&- & &271159.0& &2147.2&2161.0&2165.0& &96612.1&98510.0&98719.0
\end{tabular}
\end{ruledtabular}

PW : Present Work, Corr : Correlated values: $^a$Reference \cite{cle};
$^b$Reference \cite{sar}; $^c$Reference \cite{tri}; $^d$Reference \cite{tri1}
\end{table*}
 
Since the Kohn - Sham theory, in its very spirit, provides an effective local 
one- body potential in which the mutually noninteracting electrons are
immersed, the quantal exchange-correlation effects of the electron-electron
interactions (after filtering out the ``classical'' part $\int n(\vec{r'}) 
d^3 r' / \mid \vec{r} - \vec{r'} \mid$ in $v_{eff}^{KS} (\vec{r}))$ embody 
a kinetic energy like piece and a potential energy like piece. 
The kinetic piece arises out of the difference between the {\it functionals}
$T[n]$ and $T_s[n]$, whose values albeit known, their {\it forms} remain unknown.
Lam and Platzman \cite{LP} and Tong and Lam \cite{TL} imported the functional
forms for a homogeneous electron gas and estimated the correction to $J(q)$
within the local density approximation. In the present case however, the 
difference as a {\it functional} cannot be isolated from the sum total Kohn-Sham
exchange-correlation energy.

To conclude, it is gratifying that the present method offers a general 
prescription to estimate quantal momentum space properties starting 
from coordinate space $n(\vec{r})$ alone, with no reference to the many-
electron wave function. This scheme could also be extended to solids
wherein directional Compton profiles could be derived from an experimental
\cite{jay}
three-dimensional co-ordinate space density. Note that this
procedure is not limited by the form of the external binding potential
endowing the present scheme with generality for any bound state problem.

\begin{acknowledgments}
    RKP is indebted to AS-ICTP, where a part of the work was carried out, for
an associateship.
AK wishes to acknowledge UGC (Govt. of India) for financial support.
\end{acknowledgments}


\begin{thebibliography}{99}
\bibliographystyle{unrst}

 
\bibitem{march}
See, for example, N. H. March, {\it { Self-consistent Fields in Atoms}},
(Pergamon, New York, 1975).
 
\bibitem{wil}
An excellent compendium on this is {\it {Compton Scattering}}, edited by B. G.
Williams, (McGraw Hill, Great Britain, 1977).
 
\bibitem{PT}
P. M. Platzman and N. Tozar, in Ref. \cite{wil}, p. 28; M.J. Cooper, Radia. 
Phys.  Chem. {\bf 50}, 63 (1997).
 
\bibitem{HK}
P. Hohenberg and W. Kohn, Phys. Rev. {\bf 136}, B864 (1964); W. Kohn and L.J. 
Sham, Phys. Rev. {\bf 140}, A1133 (1965); {\it{Density Functional Methods in 
Physics}}, edited by R. M. Dreizler and J. da Provid\^{e}ncia, (Plenum Press, 
New York, 1985).
 
\bibitem{bur}
G. Burkhardt, Ann. Phys. (Leipzig) {\bf 26}, 567 (1936).
 
\bibitem{kon}
A. K\`{o}nya, Hung. Acta. Phys. {\bf 1}, 12 (1949).
 
\bibitem{CM}                                  
C. A. Coulson and N. H. March, Proc. Roy. Soc. (London) {\bf 63A}, 367 (1950).
 
\bibitem{PG}
R. K. Pathak and S.R. Gadre, J. Chem. Phys. {\bf 74}, 5925 (1981); S. R. Gadre
and R. K. Pathak, Phys. Rev. A {\bf 24}, 2906 (1981); R. K. Pathak, P. V. Panat
and S. R. Gadre, Phys. Rev. A {\bf 25}, 3073 (1982); R. K. Pathak, S. P. Gejji 
and S. R. Gadre, Phys. Rev. A {\bf 29}, 3402 (1984).

\bibitem{wig}
E. P. Wigner, Phys. Rev. {\bf 40}, 749 (1932).
 
\bibitem{hil}
M. Hillery, R. F. O'Connell, M. O. Scully and E. P. Wigner, Phys. Res. Reports 
{\bf 106}, 123 (1984).
 
\bibitem{KN}
Y. S. Kim and M.E. Noz, {\it {Phase Space Picture of Quantum Mechanics}} (World
Scientific, Singapore, 1991); W. P. Schleich, {\it { Quantum optics in the Phase
Space}} (Wiley - VCH, Weinheim, 2001).
 
\bibitem{parr}
R.G. Parr, K. Rupnik and S.K. Ghosh, Phys. Rev. Lett. {\bf 56}, 1555 (1986).
 
\bibitem{ZP}
Q. Zhao and R.G. Parr, Phys. Rev. A {\bf 46}, 2337 (1992); J.Chem. Phys. 
{\bf 98}, 543 (1993).
 
\bibitem{Levy}
M.Levy, Proc. Natl. Acad. Sci. (USA) {\bf 76}, 6062 (1978).
 
\bibitem{mor}
R. Morrison and Q. Zhao, Phys. Rev. A {\bf 51}, 1980 (1995).

\bibitem{ing}
V. E. Ingamells and N. C. Handy, Chem. Phys. Lett. {\bf 248}, 373 (1996).
 
\bibitem{har}
M. K. Harbola, Phys. Rev. A {\bf 69}, 042512 (2004). 
 
\bibitem{kog}
T. Koga, Y. Kasai and A.J. Thakkar, Int. J.  Quantum Chem. {\bf 46}, 
689 (1993).   

\bibitem{esq}
R.O. Esquivel and A.V. Bunge, Int. J. Quantum Chem. {\bf 32}, 295 (1987).                                                

\bibitem{bun}
A.V. Bunge and R.O. Esquivel, Phys. Rev. A {\bf 34}, 853 (1986).                

\bibitem{coo}
A.S. Coolidge and H.M. James, Phys. Rev. {\bf 49}, 676 (1936).                                        

\bibitem{GCP}
S. R. Gadre, S. Chakravorty and R. K. Pathak, J. Chem. Phys. {\bf 78}, 4581 (1983).                                        


\bibitem{cle}
E. Clementi and C. Roetti, At. Data Nucl. Data Tables {\bf 14}, 177 (1974);
 R. R. Zope, M. K. Harbola, and R. K. Pathak, Eur. Phys. J. D 7, 151 (1999).
 
\bibitem{eis}
P. Eisenberger and W. A. Reed, Phys. Rev. A {\bf 5}, 2055 (1972).

%\bibitem{smi}
%R. Benesch and V. H. Smith Jr., Phys. Rev. A {\bf 5}, 114 (1972).

\bibitem{tri}
A.N. Tripathi, R.P. Sagar, R.O. Esquivel and V.H. Smith Jr., Phys. Rev. A {\bf
45}, 4385 (1992).
 
\bibitem{TL}
B. Y. Tong and L. Lam, Phys. Rev. B {\bf 18}, 552 (1978).
 
\bibitem{eis1}
P. Eisenberger, Phys. Rev. A {\bf 5}, 628 (1972).
 
\bibitem{sar}
A. Sarsa, F.J. G\'{a}lvez and E. Buendia, J. Phys. B : At. Mol. Op. Phys. 
{\bf 32}, 2245 (1999).
 
\bibitem{tri1}
A.N. Tripathi, V.H. Smith Jr., R.P. Sagar and R.O. Esquivel, Phys. Rev. A
{\bf 54}, 1877 (1996).

\bibitem{LP}
L.Lam and P. M. Platzman, Phys. Rev. B {\bf 9}, 5122 (1974); {\it ibid} B 
{\bf 9}, 5128 (1974).


\bibitem{jay}
D. Jayatilaka, Phys. Rev. Lett., {\bf 80}, 798 (1998).  
\end{thebibliography}
\end{document}